\documentclass[conference]{IEEEtran}
\IEEEoverridecommandlockouts
\usepackage{cite}
\usepackage{amsmath,amssymb,amsfonts}
\usepackage{algorithmic}
\usepackage{graphicx}
\usepackage{textcomp}
\usepackage{xcolor}
\usepackage{hyperref}

\usepackage[numbers]{natbib}
\def\BibTeX{{\rm B\kern-.05em{\sc i\kern-.025em b}\kern-.08em
    T\kern-.1667em\lower.7ex\hbox{E}\kern-.125emX}}
\begin{document}

\title{Evaluating Digital Library Search Systems by using Formal Process Modelling}

\author{\IEEEauthorblockN{Christin Katharina Kreutz\IEEEauthorrefmark{1}, Martin Blum\IEEEauthorrefmark{2}, Philipp Schaer\IEEEauthorrefmark{1}, Ralf Schenkel\IEEEauthorrefmark{3}, Benjamin Weyers\IEEEauthorrefmark{3}}
\IEEEauthorblockA{\IEEEauthorrefmark{1}\textit{TH Köln - University of Applied Sciences}, Cologne, Germany \\
\IEEEauthorrefmark{2}\textit{Schloss Dagstuhl LZI, dblp}, Trier, Germany \\
\IEEEauthorrefmark{3}\textit{Trier University}, Trier, Germany \\
\IEEEauthorrefmark{1}\{christin.kreutz, philipp.schaer\}@th-koeln.de, \IEEEauthorrefmark{2}martin.blum@dagstuhl.de, \IEEEauthorrefmark{3}\{schenkel, weyers\}@uni-trier.de\\
}
}

\maketitle

\begin{abstract}
Evaluations of digital library information systems are typically centred on users correctly, efficiently, and quickly performing predefined tasks. Additionally, users generally enjoy working with the evaluated system, and completed questionnaires show an interface's excellent user experience. 
However, such evaluations do not explicitly consider comparing or connecting user-specific information-seeking behaviour with digital library system capabilities and thus overlook actual user needs or further system requirements. 

We aim to close this gap by introducing the usage of formalisations of users' task conduction strategies to compare their information needs with the capabilities of such information systems.
%
We observe users' strategies in scope of expert finding and paper search.
We propose and investigate using the business process model notation to formalise task conduction strategies and the SchenQL digital library interface as an example system. 
We conduct interviews in a qualitative evaluation with 13 participants from various backgrounds from which we derive models.

We discovered that the formalisations are suitable and helpful to mirror the strategies back to users and to compare users' ideal task conductions with capabilities of information systems.
%
We conclude using formal models for qualitative digital library studies being a suitable mean to identify current limitations and depict users' task conduction strategies.
Our published dataset containing the evaluation data can be reused to investigate other digital library systems' fit for depicting users' ideal task solutions.
\end{abstract}


\section{Introduction}

Nowadays, there are numerous options for obtaining bibliographic information. Classic bibliographic digital libraries (DLs) such as the ACM DL, 
Bibsonomy, 
dblp, 
Google Scholar, 
Semantic Scholar, 
SpringerLink, 
or ResearchGate 
support search and exploration functionality. More complex information needs that do not solely require using keyword-based search can be assessed by more sophisticated information systems. One option is to write Cypher queries in the specialised tool GrapAL~\cite{DBLP:conf/acl/BettsPA19} or to formulate RDF-style triples in the Narrative Service of PubPharm~\cite{DBLP:conf/jcdl/KrollPPB22}. Another could be using GUI-assisted query construction such as SchenQL~\cite{DBLP:conf/jcdl/KreutzBS22}.

Evaluations of such DL systems are crucial for their development as they help identify current shortcomings and room for improvement~\cite{Xie}. 
Historically, library and information science focused more on systems than on users' perspectives~\cite{seekingmeaning}. 
Nowadays, user studies are conducted in many forms with a plethora of different measures~\cite{Xie} in the domain of bibliographic metadata: 
Many approaches assess quantifiable measures such as the correctness of user-constructed queries to fulfil specific tasks~\cite{
schenqljournal,Dalkiran,Zhu2}%
, time spent to satisfy an information need~\cite{
schenqljournal,Dalkiran,Zhu2}%
, the constructed query size~\cite{Zhu2,Dinet2004SearchingFI}, the number of clicks to find the solution for a task~\cite{Dalkiran} or questionnaires with a focus on deriving subjective user feedback~\cite{
schenqljournal,Dalkiran,finna} 
 which may contain quantitative scales, such as Likert scales. Some DLs are evaluated more qualitatively by using think-aloud protocols~\cite{
Dalkiran,daffodil,DBLP:journals/jodl/KramerPCKM21}%
, query log analysis~\cite{Dinet2004SearchingFI}, open-ended questions for users~\cite{
nonusers,3diglib,Liang,Bartalesi} 
or interviews with domain experts~\cite{MARCHIONINI1998535,schenqljournal,DBLP:conf/acl/BettsPA19}. 

However, Kuhlthau~\cite{seekingmeaning} argued for the need to connect users' information-seeking behaviour and the systems providing the information, to think beyond precision-based evaluations and include users' perspectives in the search process. This corresponds to Kelly's~\cite{DBLP:journals/ftir/Kelly09} definition of \textit{Information-Seeking Behavior with IR Systems} as human-focused studies investigating users' usual information-seeking behaviour while interacting with an information system. Her mentioned prototypical studies of this category~\cite{kellar,ford,kimallen,bystroem} all produce quantifiable data that can be statistically analysed.


Overall, current DL evaluation designs and studies appear to be primarily concerned with quantitative measures, with the occasional report of anecdotal insight gained from free-text questions or (semi-structured) interviews. 
Considering this quantitative data may give hints to which elements of a DL may generate specific issues reducing search performance, it lacks of giving clear directions of further improvements. 
In this case, qualitative data with a broader scope could give better insight into the actual user needs as well as further indications of potential future developments. 
This is also the primary function of qualitative methods in designing interactive systems~\cite{rogers2011interaction}. 
A crucial factor for the design of interactive systems, including DLs, is the actual task conducted by the user with the system~\cite{bowen2021task}.
With this work at hand, we aim to consider the actual task conduction model of such search tasks conducted in DLs as the central element for the successful design and evaluation of DLs in this context~\cite{Xie}. 
Still, despite studies observing usage models~\cite{artist}, a specific qualitative evaluation of discrepancies between users' ideal task execution strategies and a systems' design has been disregarded so far.

The lack of these studies might be attributed to difficulties in formalising information exploration processes with standard process modelling tools~\cite{Tibau}.
Exploratory search is frequently preceded by an anomalous state of knowing where a searcher requires information to proceed in their task~\cite{Vakkari}. 
So the formulation of information needs is an iterative process in which new information influences a user's perception of the information space~\cite{Vakkari}.
Quantifying these changes in a user's task conduction model could assess changes in the information formulation~\cite{Vakkari} as a fine-grained micro-observation.
Contrasting this viewpoint, our macro-objective is capturing the difference in the overall representation of the task execution process of users and the functionality offered by an information system. 
We do not consider single conceptual changes of the users' task conduction models during information exploration. Instead, we investigate the users' task conduction models of the complete task in a cumulative, outcome-oriented form.

Therefore, in this work, we formalise the task conduction models derived from users' typical task completion strategies 
to compare and connect them with the capabilities of an information system. 
We propose investigating how users' task conduction models are translated to process documentations and adapted to fit the limited options of a single system with the following general research question: \textit{How can we compare users' conceptions of search tasks in digital library with capabilities of such a system?} 
As a goal, we seek to help identify dissonances in an interface's functionality that could be modified to suit the users’ pre-existing conceptions of typical tasks in such systems better

As an exemplary application of the proposed method, we use the SchenQL ecosystem~\cite{DBLP:conf/jcdl/KreutzBS22} in our study. At the time of writing this paper, this is the most recent digital library interface suitable to satisfy many complex information needs, surpassing the scope of other current DLs~\cite{DBLP:conf/jcdl/KreutzBS22}. SchenQL's usage is easy to learn, and the system is appropriate for users of DLs with different expertise levels~\cite{schenqljournal}.
We use Law et al.'s~\cite{weyers} BPMN variant to represent users' task conduction models, as this method does not incorporate a modellers perspective but instead solely focuses on capturing a user's perspective collected through interview data. 
In our \textit{Information-Seeking Behavior with IR Systems} type study (i.e., ~\cite{Kelly}), we analyse and compare BPMNs of users' ideal task conduction~\cite{weyers}, models describing users' actual task conduction using a specific system (we use SchenQL~\cite{DBLP:conf/jcdl/KreutzBS22} as an example for this work) and models translating the initial system-independent strategy to the system (see Figure~\ref{fig:mms}).

Our contribution can be summed up as follows:
\begin{itemize}
    \item Application of a formal representation method of user processes~\cite{weyers} into a digital library evaluation setup with a specific focus on identifying current shortcomings. 
    \item Description and analysis of users' ideal task conduction models for everyday tasks in DLs as well as discrepancies and adaptation when confronted with a specific DL.
    \item Exemplary analysis of our qualitative evaluation method on the SchenQL query language and user interface.
    \item Publication of reusable interviews and formal models of users' processes of solving typical tasks in DLs~\cite{DATASET}.
\end{itemize}


\section{Related Work}

We discuss areas adjacent to this work: We present \textit{bibliographic digital libraries} before compiling strategies for the \textit{evaluation of bibliographic digital libraries} and presenting different types of \textit{modelling users' task conduction}.  

\subsection{Bibliographic Digital Libraries}

The ACM DL
, Bibsonomy
~\cite{DBLP:series/xmedia/HothoJBGKSS09}, dblp
~\cite{DBLP:journals/pvldb/Ley09}, Google Scholar, 
Semantic Scholar
~\cite{s2}, Springer Link, 
ResearchGate, 
Clarivate/Web of Science~\cite{salisbury}
, Elsevier Scopus~\cite{salisbury},  
and Dimensions
~\cite{DBLP:journals/qss/HerzogHK20} are DLs operating on bibliographic metadata which offer keyword-based search to retrieve and explore the underlying data.
Some systems in the domain of bibliographic metadata provide more functionality or information than simple keyword-based search:
OpenAlex
~\cite{openalex} provides a web API and database snapshot, which allows the formulation of complex queries.
Semantic Scholar~\cite{s2} extracts paper content information, such as entities and mentions, which are used to construct a literature graph.
Clarivate/Web of Science~\cite{salisbury} offers a sophisticated faceted search interface to construct queries.
GrapAL~\cite{DBLP:conf/acl/BettsPA19} is a Cypher-based query formulation tool and GUI offering complex graph operations.
Daffodil~\cite{daffodil} is a faceted digital library interface that supports users in expressing their complex information needs by suggesting query components and marking potential errors. 
SchenQL~\cite{DBLP:conf/jcdl/KreutzBS22} (see Sec.~\ref{sec:schenQL}) is a domain-specific query language and GUI supporting information search and exploration by providing aggregations and expert functions.
%


\subsection{Evaluation of Bibliographic DLs}
Opposing our qualitative evaluation of a DL, multiple \textit{quantitative} study designs have been conducted, mainly focusing on users solving predefined tasks~\cite{Zhu2,schenqljournal,Dalkiran,Dinet2004SearchingFI,daffodil}. 
Quantitative measures such as answering time~\cite{Zhu2,schenqljournal}, correctness~\cite{Zhu2,schenqljournal}, query size~\cite{Zhu2,Dinet2004SearchingFI}, used query components~\cite{Dinet2004SearchingFI}, perceived query difficulty~\cite{schenqljournal} and query execution time~\cite{schenqljournal} were reported.





There are also some \textit{less structured evaluations} focusing on finding challenges of current systems, and users' information needs through expert interviews~\cite{schenqljournal,DBLP:conf/acl/BettsPA19,MARCHIONINI1998535}, usage diaries~\cite{artist}, questionnaires~\cite{Liang,nonusers,Bartalesi,finna,artist,DBLP:journals/jodl/KramerPCKM21}, user assessment of prototypical interfaces~\cite{MARCHIONINI1998535,daffodil}, thinking aloud interviews while conducting tasks~\cite{daffodil} and screen captured task conduction which was later annotated by users~\cite{DBLP:journals/jodl/KramerPCKM21}. 
While these methods targeted one of the goals of our study, they did not systematically connect users' perspectives with the respective systems. 
From thinking-aloud tests~\cite{daffodil}, formal cognitive models could have been derived. However, a systematic analysis of the task conduction and a comparison using the participant's usual DLs was disregarded, compared to our approach. Furthermore, when evaluating usage diaries~\cite{artist}, no tasks were predefined, contrasting our approach; another difference is disregarding the user's typical task conduction strategy without the system at hand.
Another evaluation~\cite{DBLP:journals/jodl/KramerPCKM21} only observed the typical task execution behaviour of participants and constructed formal models from the transcribed data but did not relate it back at a single system.

\subsection{Modelling Users' Task Conduction}

Task modelling has been developed in the context of engineering of interactive systems. This family of model-based approaches aims at informing and driving the design and development of interactive systems
~\cite{bowen2021task}. Either user or system prospective could be taken where the user perspective is relevant for our work. The driving question here is how a user performs a search task in a given DL such that task models enable to describe these processes and reuse them in terms of creation of DL interactive interfaces. 

Various modelling approaches and languages have been proposed in the area of engineering of interactive systems. 
ConcurTaskTrees (CTT)~\cite{paterno} is an approach to model all actions users partake to achieve specific goals. The notation can be used to describe how tasks could be solved in an existing or envisioned system or how users think a task should be performed. CTT focuses on activities, has a hierarchical structure, a graphical syntax and provides numerous temporal operators.
HAMSTERS~\cite{hamsters} is a notation based on and compatible with CTT used to decompose complex real-life models into smaller task-based, interconnected ones. Task models can communicate with each other. The notation distinguishes between task types such as system, users (e.g., cognitive or motor tasks) and interactive tasks (e.g., input or output tasks). Temporal relationships between tasks are important.

Dias et al.~\cite{Tibau} 
present the Exploratory Search KiP model to visualise information-seeking in the web. They model four activities: Search term selection, query formulation, result check and information extraction.
From click tables and explanations for clicks derived from thinking-aloud protocols, these activities are defined~\cite{TibauMT}. The modeller is explicitly involved in the design process by incorporating their interpretation of users' actions and statements to identify mismatches.

Still, literature around the creation of task models mostly neglects the actual creation process as has been identified by Bowen et al.~\cite{bowen2021task}. With the Business Process Modeling Notation (BPMN)~\cite{bpmn}, processes can be formalised in detail. Processes can be depicted through sequences of activities, conditions, paths and logical operators. The variant introduced by Law et al.~\cite{weyers} consists of a subset of the BPMN modelling options for depicting structured user interaction models from thinking-aloud interviews. They proposed a methodology to gather BPMN models in a structured way from an informal data base. These have been gathered from so called think aloud interviews, a qualitative method combining think aloud protocol with open interview techniques (see Section~\ref{sec:BPMN}). 

\section{Methodological Basics}

This section describes the combined methods in our evaluation, briefly introduced in the previous section. 

\subsection{Modelling Users' Task Conduction}
\label{sec:BPMN}

Law et al.~\cite{weyers} describe a sequential and imperative method to construct structured user interaction models from unstructured, process-oriented thinking-aloud interviews. They create models in BPMN by segmenting transcribed interviews into tasks and later classifying these segments into six categories. The process focuses on events, tasks and conditions as provided by BPMN. Artefacts such as tools or data are not specifically modelled.

Segments are created by cutting the transcripts by verbs, as they signal tasks or events. For adjective or relative clauses, the segments should not be separated. Time-related phrases should be separated, even without the presence of verbs. In general, a finer segmentation is preferred. 

\begin{figure*}
    \centering
    \includegraphics[width=0.75\textwidth]{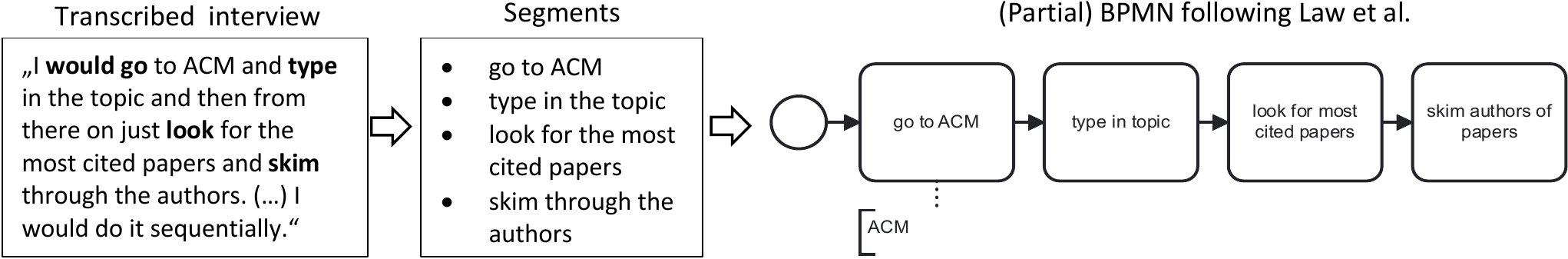}
    \caption{Construction of the (partial and unverified) BPMN from transcribed interview data for participant \textit{blue\_dog}.}
    \label{fig:bpmn}
\end{figure*}

In the following classification step, segments are identified to be either of type \texttt{setting}, \texttt{annotation}, \texttt{task}, \texttt{event}, \texttt{condition} or \texttt{other}. In our use case, the most prominent classes are tasks (activities performed by users; yellow parts in Fig.~\ref{fig:PCM_gd}), events (events occurring during task conduction, incidents in the environment; red parts in Fig.~\ref{fig:PCM_gd}), conditions (descriptions of alternatives; blue part in Fig.~\ref{fig:vIMM_gd}) and annotations (mentioned DLs, systems, observations from screen captured data; green part in Fig.~\ref{fig:vIMM_gd}). These classes correspond to specific BPMN elements.
Figure~\ref{fig:bpmn} shows an example transformation of an interview to a (partial) BPMN.

This method does not focus on the details of activities but rather on switching tasks. Its structure aims at preventing including the modeller's interpretation of the interviews and focuses on the interview content. This has been demonstrated in the original article by an empirical evaluation of the method. 

\subsection{The SchenQL Query Language and GUI}
\label{sec:schenQL}

\begin{figure*}
\centering
    \includegraphics[width=0.75\textwidth]{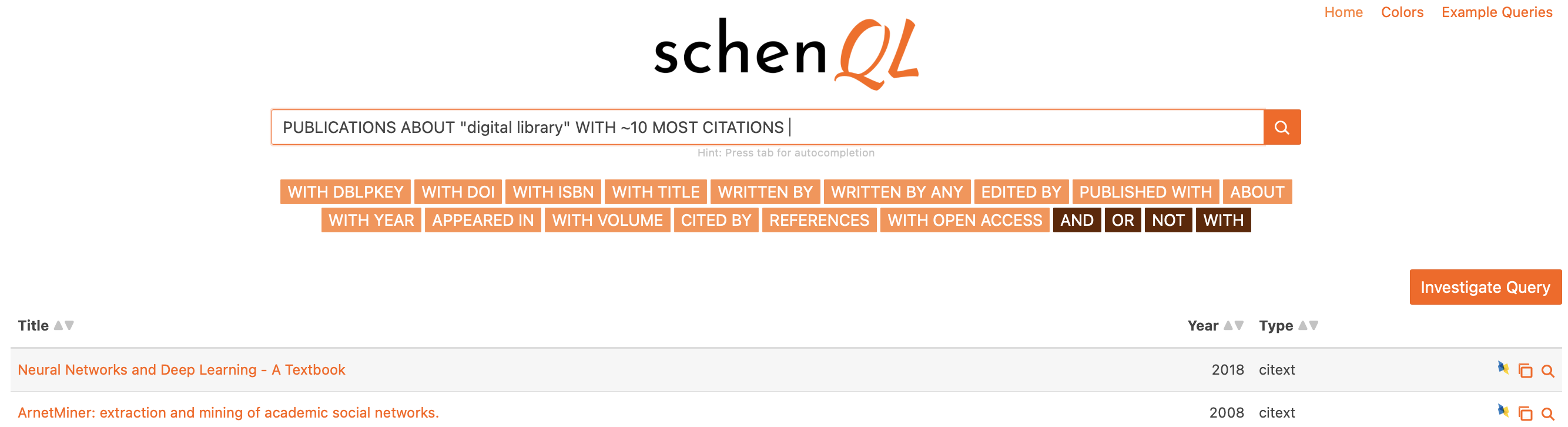}
    \caption{SchenQL search interface with colour-coded (dependent on the component type) query component suggestions.}
    \label{fig:search}
\end{figure*}

SchenQL~\cite{DBLP:conf/jcdl/KreutzBS22} is a domain-specific query language and graphical user interface (GUI) supporting information search and exploration. Its main goal is to support various types of users of digital libraries in their information needs. It offers a broad selection of domain-specific functions such as co-author search, citation aggregation or providing bibliographic metrics.
The system offers query formulation following a predefined sophisticated grammar as well as information exploration via the interface. Its GUI offers suggestions and auto-completion of query components to help users with query formulation (shown in Figure~\ref{fig:search}).
SchenQL was designed with the dblp computer science bibliography as the central use case.
Kreutz et al.~\cite{DBLP:conf/jcdl/KreutzBS22} give an overview of SchenQL's grammar and GUI.

We selected SchenQL as an example DL in our studies for its recency, meaning the system is still being in use or updated~\cite{DBLP:conf/jcdl/KreutzBS22}, its suitability for users of different experience with DLs~\cite{schenqljournal}, its ease of use~\cite{schenqljournal} and mainly for its broad coverage of bibliographic information needs, surpassing the scope of most current 
DLs~\cite{DBLP:conf/jcdl/KreutzBS22}.

\subsection{General Idea}

\begin{figure*}
    \centering    
    \begin{minipage}[t]{0.65\textwidth}
    \includegraphics[width=\textwidth]{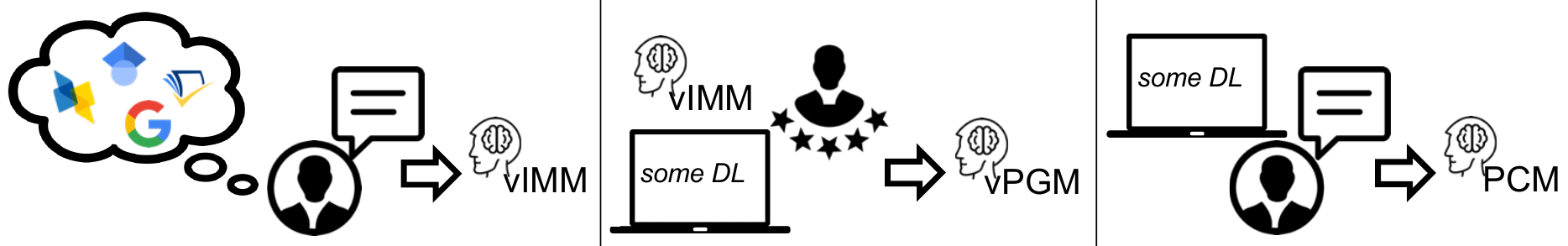}
    \end{minipage}
    \hfill
    \begin{minipage}[t]{0.34\textwidth}

    \vspace{-57.5pt}
\scriptsize
\begin{tabular}{l|p{4.6cm}}
        model & explanation \\\hline
        vIMM & user's general or ideal strategy to solve a task using their usually preferred systems\\
        vPGM & translation of a user's ideal task solution to the DL which is evaluated by an expert on the system\\
        PCM & user's actual strategy solving the task by using the specific DL which is evaluated\\
    \end{tabular}
    
    \end{minipage}

    \caption{Construction of the three task conduction models: The vIMM comes from a participant's general usage of digital libraries or other sources, the vPGM comes from an expert's translation of the vIMM to the DL which is being evaluated and the PCM comes from the participant using that system.}
    \label{fig:mms}
\end{figure*}

Our general idea is to use users' usual process for task conduction and their solution strategy when using a specific digital library system to evaluate DL systems. For this, we compose three models for each task and user:
\textit{First} we construct a BPMN from a study participant's interview on their general process for solving a task using their preferred digital libraries following the method by Law et al.~\cite{weyers}. Then, the same user verifies their ideal task conduction model, which results in their \textbf{v}erified \textbf{I}nterview \textbf{M}ental \textbf{M}odel (\textbf{vIMM}). 
The \textit{second} model we construct (the \textbf{v}erified \textbf{P}rocess \textbf{G}old \textbf{M}odel, \textbf{vPGM}) is an expert-generated translation of a user's workflow from their chosen DLs and tools to the example digital library system. This BPMN represents the application of a user's strategy as closely as possible to the DL being evaluated.
The \textit{third} model (the \textbf{P}rocess \textbf{C}onduction \textbf{M}odel, \textbf{PCM}) is composed by a user's actual execution of the task with the DL, which is being evaluated. This BPMN represents users' actual task solution process using a specific system. 

Figure~\ref{fig:mms} shows a simplified construction of the three models.
In this work, we choose SchenQL~\cite{DBLP:conf/jcdl/KreutzBS22} as an exemplary DL to demonstrate the evaluation process and analyse the results.


\section{Experimental Setup}

This section presents our research questions, tasks, the study participants, 
our evaluation setup, and technical details.

We compare the model of users' general task conduction produced with Law et al.'s~\cite{weyers} method with a model composed from users' task conduction in a specific system (SchenQL~\cite{DBLP:conf/jcdl/KreutzBS22}) and a model of the translation of their initial system-independent strategy to the system (see Figure~\ref{fig:mms}).

\subsection{Research Questions}
\label{rqs}

We observe different aspects related to our general question \textit{How can we compare users' conceptions of search tasks in a digital library with capabilities of such a system?}
We investigate the following five more fine-grained research questions by using Law et al.'s~\cite{weyers} method to represent users' task conduction models and SchenQL~\cite{DBLP:conf/jcdl/KreutzBS22} as the example DL system for this work to observe the suitability and use of formalised task conduction models:

\begin{itemize}
    \item[RQ$_1$] What are users' preferences, which components of digital libraries are usually used for the 
    predefined tasks?
    \item[RQ$_2$] How do users utilise the example system, which components are used for the specific predefined tasks?
    \item[RQ$_3$] What are the limitations of the example DL system? Which components or functions were ignored or missed?
    \item[RQ$_4$] Is the example system usable for advanced DL tasks?

    \item[RQ$_5$] What are the discrepancies between the ideal task conduction models of users and their actual task conduction, how are models adapted to solve the predefined tasks?
\end{itemize}

\subsection{Tasks}


We focus on two deliberately vague exploratory task descriptions, giving study participants some wiggle room to fit their usual information-seeking behaviour better. For instance, a user might define \textit{expertise} as having published many papers in specific topic-related journals, while another might consider those with a high topic-independent $h$ index as relevant.

\begin{itemize}
    \item Task \textbf{T$_{ex}$}: Find two experts on a topic of your liking. 
    \item Task \textbf{T$_{pa}$}: Find relevant papers from a topic of your liking which appeared after 2017.\footnote{Soufan et al.~\cite{DBLP:conf/chiir/SoufanRA22} provide an overview on the exploratory search task.} 
\end{itemize}

\subsection{Participants}

Our thirteen study participants are computer or information scientists with differing expertise in using DLs for research tasks, by which they were invited to take part
: Two masters students, six PhD students (first year to last year students), an industry researcher, a dblp staff member, a postdoc and two professors. 
They participated voluntarily and did not receive any incentives. 
For anonymisation, participants chose code names with which we refer to them.

\subsection{Setup and Experiment Conduction}

        


Our experimental setup is composed of four consecutive parts, with 
seven steps. 
Users were part of two user sessions; other parts did not include the study participants. 
They were evaluated on a one-by-one basis with one (the same) investigator present.
In the following, the seven steps are described.

\textbf{Step \textit{i)}: Pre-Interview Questionnaire.}
Step \textit{i)} and \textit{ii)} were conducted sequentially and took about 30 minutes. They form the first user session.
Participants filled out a questionnaire to consent to voluntarily participate in the study, have their voices and screens recorded, and have transcripts as well as formalised task conduction models published. 
All participants were explicitly made aware that they could stop and drop out of the experiment at any time without consequences. 
The process (a mail to the investigator) for later deleting user-specific data from the resulting dataset was also explained.
The conduction of an initial questionnaire was accompanied by an interviewer and audio recorded.


\textbf{Step \textit{ii)}: Interview.}
In recorded semi-structured 1:1 interviews with the participants, they described how they usually conducted tasks T$_{ex}$ and T$_{pa}$ with a focus on the data sources.
For T$_{ex}$, the participants' topics, their familiarity with the topics, their definition of an expert and their process of solving T$_{ex}$ were asked. 
Afterwards, the participants chose a topic for the second task, indicated their expertise in this topic, and gave their notion of relevancy before they described their process to solve T$_{pa}$.
The interviewer did not intervene with the participants' understanding of the tasks but posed clarifying questions regarding temporal or process-related ambiguities. 

\textbf{Step \textit{iii)}: Modelling I.}
From transcripts of the interviews for participants, one modeller constructed mental models, so-called \textit{Interview Mental Models (IMMs)}, for all tasks and users. The formation of IMMs with BPMN followed the description by Law et al.~\cite{weyers} with mentioned information sources such as digital libraries included in annotations. 

\textbf{Step \textit{iv)}: Verification.}
Step \textit{iv)} to \textit{vi)} (the second user session) 
were done in one session and took about one hour. 
%
This step verified the modeller-generated IMMs by the participant. Finally, the investigator went through the IMMs for the two tasks and asked the participant to intervene and note changes if the formalisation did not depict that user's usual process. 

\textbf{Step \textit{v)}: Tasks.}
Participants watched the five-minute video demonstration\footnote{\url{https://youtu.be/pkaKe7vo9ys}} of SchenQL by Kreutz et al.~\cite{DBLP:conf/jcdl/KreutzBS22} and had access to a language documentation. 
The screen (e.g., users' mouse movement or entered queries) and think-aloud protocol 
were recorded.
Users conducted T$_{ex}$ and then T$_{pa}$ and stopped with a task when feeling they solved it or after 25 minutes for T$_{ex}$ and 15 minutes for T$_{pa}$. 

\textbf{Step \textit{vi)}: Post-Task Questionnaire.}
 In a questionnaire, participants indicated the components that they enjoyed using in the SchenQL interface, components missing or not found and encountered problems. They could also mention anything they did not get the chance to or forgot to mention earlier.

\textbf{Step \textit{vii)}: Modelling II.}
First, the modeller revised the IMMs according to the annotations made in step \textit{iv)} to represent the user's information-seeking strategy. The resulting model is the so-called \textit{verified IMM (vIMM)}. 

Based on vIMMs, a SchenQL expert translated the workflow as closely as possible to the SchenQL system in a \textit{Process Gold Model (PGM)} in BPMN. SchenQL queries were included as annotations.
A second independent SchenQL expert who was previously uninvolved in the evaluation process 
verified the models. When the two SchenQL experts disagreed on the workflow, they discussed the issue until reaching a consensus. The \textit{verified PGMs (vPGMs)} depict the revised translation of the vIMMs to the example DL SchenQL.

For all users and tasks, a user's actual \textit{Process Conduction Model (PCM)} was constructed from their think-aloud protocol with BPMN following Law et al.~\cite{weyers} when using the exemplary system. Screen-captured data and action non-inducing verbs such as \textit{assume}, \textit{feel}\footnote{The classification of segments which describe feelings as \textit{annotation} deviates from Law et al.'s~\cite{weyers} to improve consistency of BPMNs. Such sentences normally explain user behaviour. In our case, most explanations stem from screen capture observations and are modeled as \textit{annotations}.} and \textit{wonder} were used to annotate the PCM, e.g., with constructed SchenQL queries. Non-descriptive sentences such as \textit{"oh, interesting"} while clicking a result or \textit{"okay, $h$ index of 40"} are used for further annotation. When a participant was unable to finish a task in the reserved time frame, the PGM includes the information of the task conduction stopping.

Our experiment produces three models for each user and task: A user's verified general task conduction (vIMM), a verified translation of the task to the single digital library to evaluate (vPGM), and a model of the user's actual task conduction using that DL (PCM).

\subsection{Technical Details}

As data source, we follow Kreutz et al.~\cite{DBLP:conf/jcdl/KreutzBS22} and use the dblp XML dump from 1st Oct. '21\footnote{\url{https://dblp.org/xml/release/dblp-2021-10-01.xml.gz}} and Semantic Scholar data~\cite{s2} from Oct. '21 for citations and AMiner Open Academic Graph 2.1~\cite{DBLP:conf/www/SinhaSSMEHW15,DBLP:conf/kdd/TangZYLZS08,DBLP:conf/kdd/ZhangLTDYZGWSLW19} for identifying automatically generated keywords of publications; abstracts are taken from both collections.
We used Descript\footnote{\url{https://www.descript.com}} for transcription and manually corrected the transcripts. We follow the transcription rules of Dresing and Pehl~\cite{dresing}. De-anonymising parts of the transcripts (e.g., mention of names) and the questionnaires were marked and replaced.
BPMNs were created with BPMN.io\footnote{\url{https://bpmn.io}}.

\section{Observations from Collected Data}
\label{sec:observation}

The following observations stem from the data we collected during our study via 
interviews and the constructed formal models for the two tasks of our thirteen study participants.

First, we present and discuss the three models of a specific participant as an example before the general and more fine-grained observations are laid out.

\subsection{Exemplary Data Series in Context of all Data}
\label{sec:example}

\begin{figure*}
    \centering
    \includegraphics[width=0.91\textwidth]{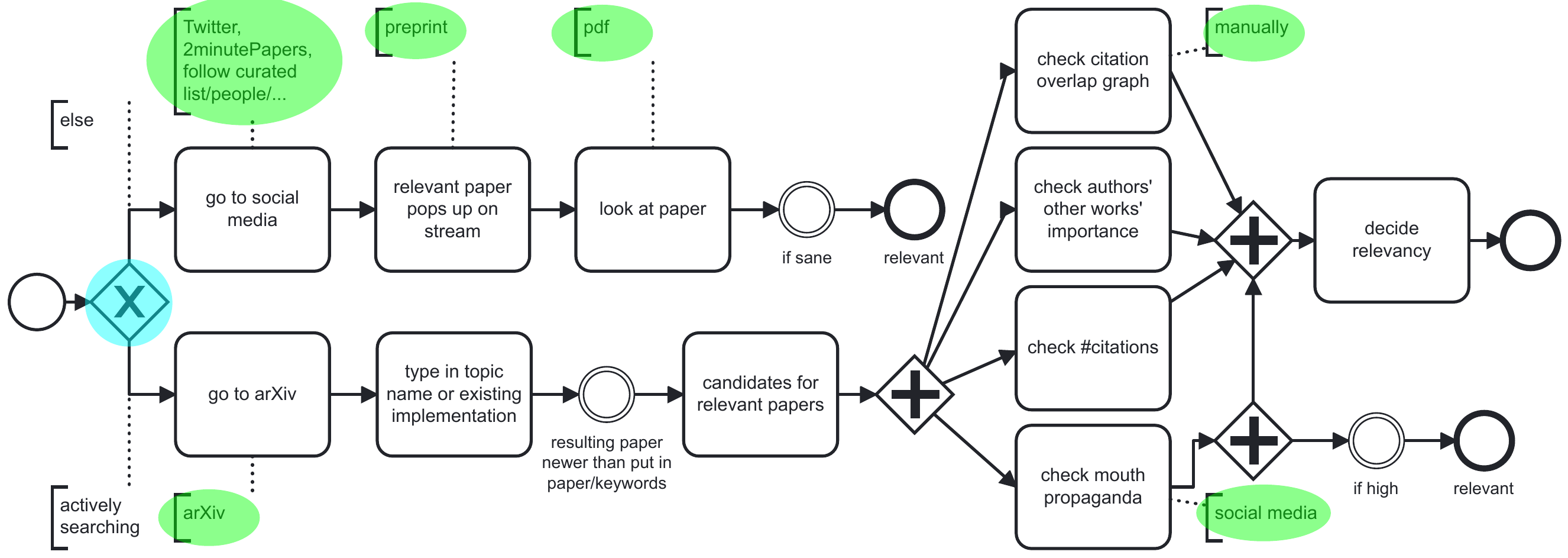}
    \caption{vIMM of T$_{pa}$ for participant \textit{green\_deer}.}
    \label{fig:vIMM_gd}
\end{figure*}

\begin{figure*}
    \centering
    \includegraphics[width=0.91\textwidth]{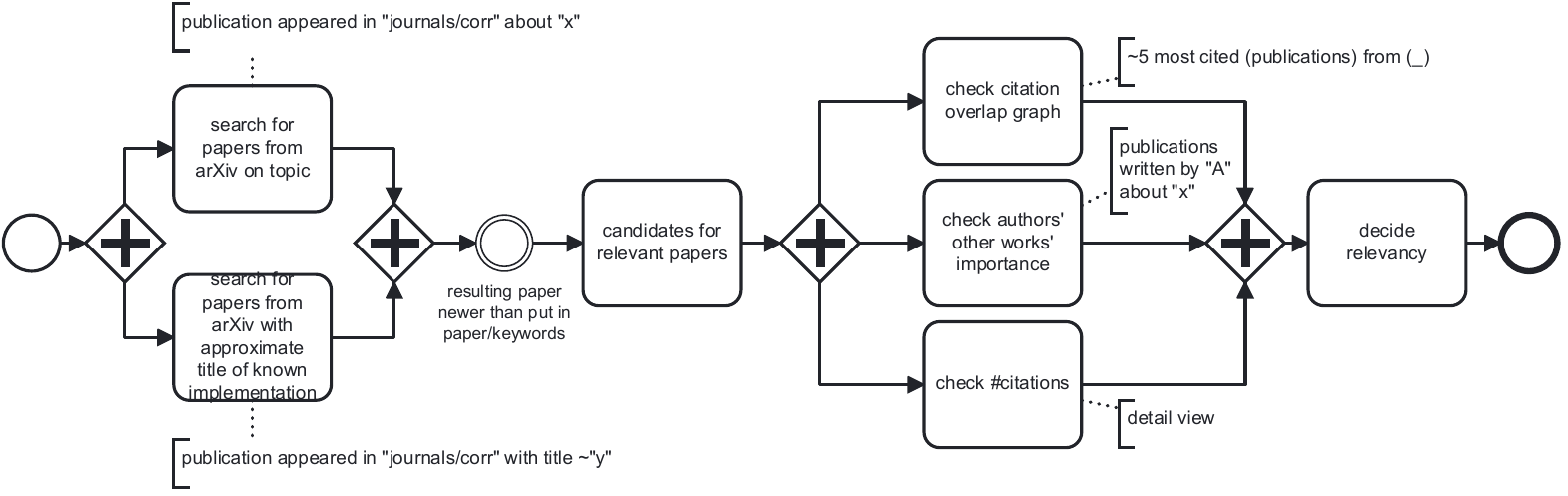}
    \caption{vPGM of T$_{pa}$ for participant \textit{green\_deer}.}
    \label{fig:vPGM_gd}
\end{figure*}

\begin{figure*}
    \centering
    \includegraphics[width=0.91\textwidth]{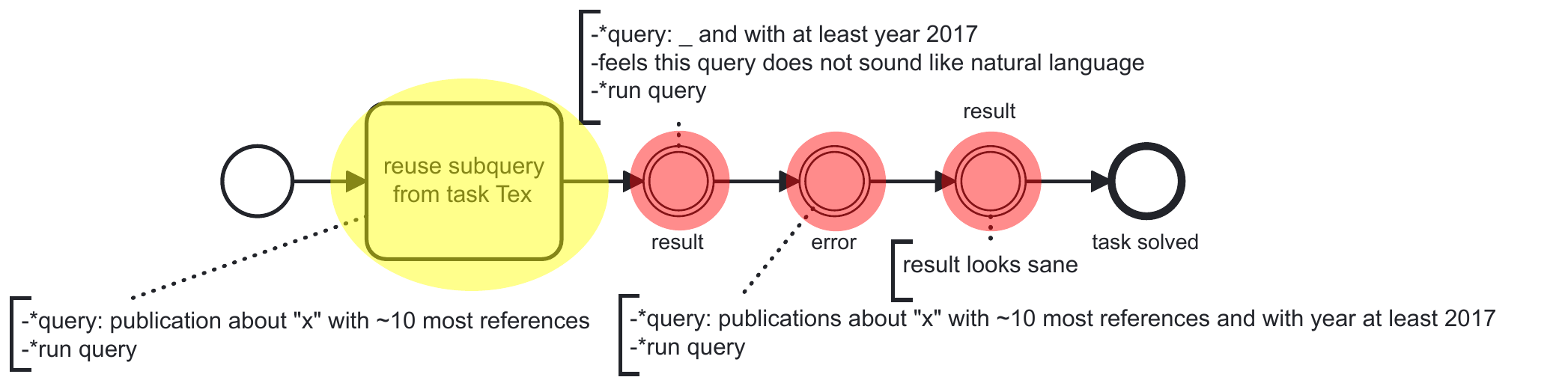}
    \caption{PCM of T$_{pa}$ for participant \textit{green\_deer} with annotations preceded by * being observations from the screen capture.}
    \label{fig:PCM_gd}
\end{figure*}

To highlight the differences in constructed models and the general reoccurring observations or problems, this section discusses the BPMNs of participant \textit{green\_deer}\footnote{The participant explicitly consented to their BPMNs being used as example models in this paper.} for T$_{pa}$ concerning all observed data. 
Figures~\ref{fig:vIMM_gd}, ~\ref{fig:vPGM_gd} and ~\ref{fig:PCM_gd} depict the vIMM, vPGM and PCM.

In general, several (4 for T$_{ex}$, 7 for T$_{pa}$) vIMMs contain multiple arms to begin with, which depend on a condition or are usually conducted in parallel (see, e.g., \textit{actively searching}, Fig.~\ref{fig:vIMM_gd}).
For some vIMMs we encountered the problem of some parts of models being very unspecific as seen with the segment \textit{check authors' other works' importance} (see Fig.~\ref{fig:vIMM_gd}). Many participants did not mention with the support of which specific system they conducted some segments (see, e.g., \textit{check \#citations}, Fig.~\ref{fig:vIMM_gd}). Several segments could not be translated to SchenQL at all, e.g., if they included outside help (see, e.g., \textit{check mouth propaganda}, Fig.~\ref{fig:vIMM_gd}), or needed to be modified in the translation process (see, e.g., \textit{go to arXiv}, Fig.~\ref{fig:vIMM_gd} translated to parallel tasks in  Fig.~\ref{fig:vPGM_gd}).
With PCMs we often faced the problem of participants not describing what they were doing or intending to do. They also rarely followed their usual strategy described in their vIMM. PCMs thus heavily rely on observations from screen capture such as clicking information or entering queries.
%
%
While users conducted tasks in SchenQL, the investigator did not interfere with their understanding of having solved the task, so the PCMs depict the user's satisfaction rather than the correctness of their result.  

More specific to the conduction models for $green\_deer$, the participant seemed to only focus on papers about their chosen topic, they did not search for (known) existing implementations on the topic (compare Fig.~\ref{fig:vIMM_gd} and~\ref{fig:PCM_gd}). The source or novelty of resulting publications also seemed to be irrelevant when using SchenQL as not only preprints were searched for (compare Fig.~\ref{fig:vIMM_gd} and~\ref{fig:PCM_gd}). This lack of restricting the result set might stem from the user's limited knowledge of SchenQL's capabilities.
Comparing the vIMM with the PCM leads to the observation that the participant only actively tries\footnote{From T$_{ex}$ we assume they confuse \textit{references} with \textit{citations}.} to incorporate one (\textit{check \#citations}, see Fig.~\ref{fig:vIMM_gd}) of the four described decision criteria. From our collected data we cannot determine if they, e.g., also included the check of authors' other works' importance by recognising resulting paper titles and remembering their authors and their full body of work. 

\textit{Green\_deer} seemed only to do part of their vIMM to solve the task. This specific part can be recognised in the PGM, where it has been fleshed out in more detail.

\subsection{Verified Interview Mental Models (vIMMs)}

\subsubsection{vIMMs for T$_{ex}$}
Popular components in solving T$_{ex}$ are keyword search (12) for papers or venues (described by Bates~\cite{Bates1989TheDO} as identifying central venues), considering authors of popular or somehow good papers as experts (7), looking at the number of citations (7), affiliations (6) and taking a deeper look into the paper, e.g., by checking the references in the related work or introduction (6). 
Five participants incorporated some undefined notion of relevancy or expertise.
Three participants searched for surveys, looked at abstracts of papers, considered author positions on papers, venue ranking, asked others for their opinion or followed references of papers (while only one participant followed a paper's incoming citations).
Related terms and query formulation or the most cited papers were important for two.
All participants, except one, usually use digital libraries to solve the expert search task.

For this task, only two participants described using a single system, nine users explicitly described a switch of their used tool throughout their process. Only two participants described the process with multiple (disjunctive) options to start solving the task. As for the system or tool to start with, multiple mentions were possible:
Four described using Google, the same amount of people started the process by using Google Scholar. 
Throughout their process, over half (7) of the participants used Google Scholar, and the same amount incorporated a Google search. These observations partially mirror users' previous indications of DLs or systems they normally use.

\subsubsection{vIMMs for T$_{pa}$}
Most participants (12) would use keyword search in T$_{pa}$.
Less than half of the participants (6) incorporate following references (described by Bates~\cite{Bates1989TheDO} as backwards chasing), using related terms or refining queries and asking other people for their opinion or expertise while searching for relevant papers.
In five cases, abstracts of papers are read or participants take a deeper dive into papers, e.g., check references in the related work section or look at figures of papers. The same amount of people use a somewhat unclear definition of relevance such as \textit{"author's other works' importance"} or \textit{"good authors"}.
The number of citations is relevant for four, venue rankings and following citations are part of three vIMMs.
Constructing a citation graph and searching for surveys were parts of two models each.

For this task, no participants described using only a single system, ten users explicitly described a switch of their used tool throughout their process.
Five participants described the process with multiple (disjunctive) options to start solving the task. As for the system or tool to start with, we did not encounter clear preferences:
Four described using Google, three people started the process by using Google Scholar and dblp. 
Throughout their process, only five participants described using Google Scholar, four mentioned using Google.
Used systems seemed to vary more compared to T$_{ex}$.

\subsection{Translation of Verified Interview Mental Models (vIMMs) into Verified Process Gold Models (vPGMs)}
\label{sec:transvIMMtovPGM}

We encountered several problems with translating vIMMs to vPGMs with our example DL. We were not able to formulate several segments of users' usual workflow throughout both tasks: 
Eight cases of participants getting help from humans in their vIMMs in T$_{ex}$,
five times someone used a general keyword-based search in Google to get an overview of their topic or trust Google's ranking, 
two users wanted to incorporate author positions in papers into their workflow, four instances occurred where persons intended to check papers related to a specific one.
Twice the restriction of papers from a specific publisher was not modelled and one participant wanted to contact persons directly.
In two cases, we encountered used data sources or information not contained in our dataset.
In three cases, we logically reordered segments in vPGMs compared to vIMMs. All these cases were related to downloads of pdfs and checking the information in these pdfs.
%
%
We could eliminate the restriction of asking others for information once, as the underlying dataset contained the data. In another case, we were able to support performing a segment for which information was missing if a user was less familiar with a topic.
%
One vIMM was not translated to a vPGM as it contained no components that could be performed in SchenQL. 

\subsection{Process Conduction Models (PCMs)}

\subsubsection{PCMs for T$_{ex}$}
All participants queried for some form of publications about their specific topic with a \texttt{publications about} type query.
Eight persons even used this query as their initial one.
Many participants (9) looked at the example queries in the GUI; five even used the documentation for an in-depth look.
Even though T$_{ex}$ is about persons, only five users checked any person's profile, while most (9) opened detail views for publications. 
Slightly more participants (6) constructed a query that asked for persons (\texttt{persons authored}). Four users looked at co-authors of people.
Abstracts of papers were read in five cases, BowTie visualisations~\cite{DBLP:conf/iknow/KhazaeiH12} were actively used slightly less (4).
As for query components, four participants used the aggregation function \texttt{with}, the same amount queried for \texttt{most cited} and three used \texttt{$\sim$RANK}.


\subsubsection{PCMs for T$_{pa}$}
For this task, participants did not need to consult the example queries (5) and documentation (4) as much as for the first one.
Almost everyone's (12) first run query is some form of \texttt{publications about a topic}. Ten persons incorporated a \texttt{with year} component in one of their queries.
Slightly fewer participants (9) took the time to check the publication detail view, only six read an abstract.
Used query components of five participants were the aggregation with \texttt{with} and \texttt{$\sim$RANK}. 
Three participants used \texttt{most cited}. 
Two persons each followed citations and references, checked the full-text of papers, specifically looked at authors of papers or re-watched part of the demo video.


\subsubsection{Limitations and Problems}
\label{sec:limitRQ3}
For T$_{ex}$ participants encountered more problems than for T$_{pa}$. The following tuples depict the number of participants with the specific problem in the task conduction, e.g., (2, 3) describes two participants having trouble in T$_{ex}$ while three users encountered this problem in conducting T$_{pa}$.
Participants encountered several errors with the SchenQL system: 
They faced errors without indicating their error, which lead to query reformulations (9, 7).
Some were met with a chip replace error\footnote{Bug, where a user clicking a suggested query chip to continue their query leads to the previous word of the query being wrongfully replaced by it.} in their query where they clicked a suggestion and a query component was replaced by the chip (4, 1), 
one person was hindered by an error in an example query. 
Some participants struggled due to the limited scope of the underlying data source (2, 1).

Non-system-caused problems that participants encountered were the following: Uncertainty about the ordering of results (7, 0), insufficient displayed information in results for queries (4, 1), confusion about the difference between full-text search and not (4, 1), complicated syntax (2, 0), confusion about the rank operation (2, 3) and confusion about the difference of terms, keywords and strings (2, 1). 
Two participants started by using the system like Google. 
The following problems occurred for single users: Uncertainty about comparing numbers of citations, struggling to filter by the number of citations, confusing citations with references, uncertainty about the difference between \texttt{most cited} and \texttt{cited by}, spelling errors in their topic, trouble distinguishing colours, confusion on when to use \texttt{and} and \texttt{with}, uncertainty which year \texttt{at least} refers to as starting year and problems with the combination of query parts.

\subsection{Post-Task Questionnaire}

\subsubsection{Disliked and Desired Components}
\label{sec:ptqRQ3}

Participants mentioned the following components which they disliked: The complicated syntax (4), non-descriptive errors (3), the auto-completion hung up (2), long loading times (2), the unclear sorting (2), the unintuitive system (1), the BowTie (1), complicated query adaption (1), need for brackets (1), the colouring (1) and the map of institutions (1).
%
%
%
Participants wished for the following components: Mouseover descriptions or previews (3), more information in the GUI (3), more example queries (2), a documentation that is better searchable (2), saving option for partial queries (1), suggestions from examples (1), dynamic linting (1) and syntax correction options (1).

\subsubsection{Liked Components}
Study participants mentioned they liked the auto-completion feature (2) of SchenQL, its query component suggestion (3) and the BowTie visualisation (3) for references and citations of, e.g., papers and persons. Several (6) mentioned the example queries to be very beneficial. 


\section{Results}
\label{results}

This section analyses the collected questionnaire data and the constructed task conduction models (see Sec.~\ref{sec:observation}) for our thirteen participants concerning the aforementioned (see Sec.~\ref{rqs}) research questions.

\subsection{RQ$_1$: User's Preferences in Digital Libraries}
\textit{RQ$_1$: What are users' preferences, which components of digital libraries are usually used for the predefined tasks?}
For this RQ, we focus on analysing the 
vIMMs concerning the conducted tasks, used systems and reoccurring patterns to find out which parts are usually used for our two tasks.

The exploration strategies for participants seem to be very diverse and consist of complex parallel steps. 
We see a tendency to start with a search engine when participants describe their usual task conduction for T$_{ex}$, for T$_{pa}$ we encounter less clear tendencies. Generally, the second task's strategies seem to rely on a more diverse set of tools.

Almost all participants described using more than one specific system and specifically incorporating a change of used tools in their process, as used systems do not seem to support all their information-seeking strategies. This might lead to switching the working sphere~\cite{switch}, which could negatively affect a user's cognitive load~\cite{cognitiveload} as well as the degree of user interaction and the systems' usability~\cite{DBLP:conf/sigir/YuanB07}. 

The constructed models for T$_{pa}$ seem to be less complex than the ones for T$_{ex}$, which could be interpreted as T$_{pa}$ being easier than T$_{ex}$. The observed need for explicitly changing tools in the second task seems to contradict this assumption.

There are numerous vaguely defined components in the vIMMs which can be interpreted diversely (e.g., \textit{check authors' other works' importance} as seen in Figure~\ref{fig:vIMM_gd}). This phenomenon is a characteristic of users' pre-focus stage~\cite{DBLP:journals/jd/VakkariH00}.

Some participants told us about another participant having been their role model/mentor in composing their exploration strategy (information on the pairs could threaten anonymity so we do not report it), but we did not see noteworthy overlap in the pairs' strategies. This leads us to assume that users of DLs adapt their usage strategy depending on their preferences.

\subsection{RQ$_2$: Used Components}
\textit{RQ$_2$: How do users utilise the example system, which components are used for the specific predefined tasks?}
In this research question, we analyse PCMs from study participants' actual task conductions with SchenQL. 

Participants of our study heavily relied on the provided example queries to start using the system.
In general, participants' tendency to explore the search space by searching for papers on a topic and deciding on experts based on their authored works became apparent. Most users stuck to comparatively simple search queries, and only few incorporated more advanced concepts such as aggregations, venue ranks or citation counts in queries. If such measures were considered, they were checked in detail views rather than queries and only in the expert search task.
The feature to investigate a query was only used once.

\subsection{RQ$_3$: System Limitations}
\textit{RQ$_3$: What are the limitations and missing components of the example system?} 
To answer this, participants' vPGMs, PCMs and post-task questionnaires were observed.

SchenQL cannot model interaction with experts, which is a characteristic of users' pre-focus stage in information search~\cite{DBLP:journals/jd/VakkariH00}, incorporation of other data sources or tools and the type of explorative search found with Google.
The errors SchenQL throws are non-descriptive, participants were uncertain about the ordering of results, they required more data to be displayed and they were confused with the differences in full-text search modes. Study participants disliked the example system's complicated syntax.
All problems are described in Sections~\ref{sec:limitRQ3} and~\ref{sec:ptqRQ3}.

\subsection{RQ$_4$: Usability for Advanced Tasks}
\textit{RQ$_4$: Is the example system usable for advanced tasks of users of DLs?}
The tasks chosen in our evaluation are ones of users of digital libraries, which can be solved by using SchenQL in theory. This RQ examines if users can satisfy their information needs even if some wished-for components they usually utilise might be missing.
%
%
All except one vIMM could be translated to vPGMs; this model for T$_{ex}$ did not contain any component which SchenQL supports. Therefore we conclude that although several modifications had to be done (see Sec.~\ref{sec:transvIMMtovPGM}), SchenQL is generally usable for conducting users' processes. 
Seven participants felt they solved T$_{ex}$, eight participants indicated they solved T$_{pa}$.

\subsection{RQ$_5$: Discrepancies and Adaption of Task Models}
\textit{RQ$_5$: What are the discrepancies between the ideal task conduction models of users and their actual task conduction, how are models adapted to solve the predefined tasks?}
Here, one DL expert and one SchenQL expert uninvolved in the model construction compare vIMMs/vPGMs and PCMs.

The experts agreed that only one participant followed their vIMM for T$_{ex}$ and two for T$_{pa}$. They found little overlap in vIMMs and PCMs.
Some participants 
seem to have forgotten their vIMM or do not know how to translate it to SchenQL while trying to resolve syntax errors, while others without many syntax problems (partially) stuck to their vIMMs. 

In the beginning, participants seemed to follow their models; then they were simplified, e.g., by disregarding decision criteria as seen with \textit{green\_deer} (see Sec.~\ref{sec:example}). This could stem from fatigue or the already retrieved information being sufficient to make decisions. 
Users' models could be strongly shaped by the available features of regularly used tools, e.g., some participants examined different quality measures than those described in their vIMM. They might not have found the usually considered one or use whatever measures a tool provides.
Our limited time frame also prevented users from entirely conducting their usual process and might have led to adaptions of conducted workflow and applied effort~\cite{DBLP:conf/chiir/CrescenziCCL21}. Instead, users focused on part of their ideal execution, possibly their models' most important parts~\cite{edlandsvenson}.

In the post-task questionnaire, one participant mentioned trying to replicate their usual process, which led to problems with getting around and getting used to the workflow with SchenQL.
Another participant voiced a similar thought in the conduction of T$_{ex}$; they checked examples for \textit{"what can I do with this tool"} instead of applying their usual strategy.

Another problem might be the participants' tendency to over-model the vIMMs. One participant answered in the post-task questionnaire \textit{"I strongly idealized my search behaviour. (...) 
my \emph{real} search behaviour is much simpler."}.

\subsection{Discussion}

\subsubsection{Usage of BPMNs}
From our sessions with participants, we observed that the constructed BPMNs were a suitable mean to discuss their task conduction strategies, that these models were generally understood and a clear way to present complex thought processes back to the user (as seen by modifications or clarifications in the verification step of our study). 
Discussions between the SchenQL experts for constructing the vPGM as well as those between the expert in digital libraries and the SchenQL expert for RQ$_5$ highlighted the used BPMNs' suitability to formalise the collected data.
Our analysis of research questions based on the constructed BPMNs further proves the approach's potential for DL systems' evaluations.

From observations related to the PCMs we found that Law et al.'s~\cite{weyers} method alone has limited suitability for modelling task conduction from think-aloud protocols if users tend to not verbalise crucial actions, even if they are regularly reminded to do so. Additional annotation data, which could be collected from eye-tracking, could help mitigate this problem.

\subsubsection{Results of the Evaluation}
When participants described their usual task conduction, they relied on multiple systems and willingly switched tools. Reasons for these switches might be habit or the tools not offering all required functionalities.
When using the example system, users would not be forced to switch tools for a large portion of their usual task conduction, as seen by the considerable part of vIMMs translated to SchenQL.
Additionally, many subsequent steps from vIMMs could have been modelled as a single step using SchenQL. This leads to the conclusion that the exemplary evaluated system SchenQL was built for one-shot queries rather than complex processes.

When participants were asked about their usual solutions for the tasks, they described complex processes, possibly caused by the Hawthorne effect~\cite{Hawthorne}, but when using a system, they show a "now or never" mentality and only conducted parts of their plan. 
As possible reasons for this, 
we assume adjustments made out of time constraints of the evaluation~\cite{DBLP:conf/chiir/CrescenziCCL21,edlandsvenson,DBLP:conf/asist/CrescenziCA13}, accessibility of features~\cite{DBLP:conf/asist/CrescenziCA13}, the off-throwing syntax errors, the unknown language/system, missing data and maybe users' information need being satisfied by using a single decision criterion so they did not need to investigate further. 
A verbalisation of a task conduction could differ from a user's mental conception~\cite{DBLP:journals/crl/Taylor15}, a BPMN depicting the verbalisation again blurs this model. The model could also change the more a participant finds out about their objective during the search~\cite{cole}.
Generally, vIMMs seemed to be over-modelled.

By the number of problems encountered in the PCMs, participants seemed to have more trouble solving T$_{ex}$ than T$_{pa}$. 
This could be attributed to learning effects or task T$_{ex}$ seemingly not to be an everyday information need for part of the participants. 
In general, the PCMs show users' tendency to iterate over queries multiple times, but only some participants consulted the documentation or example queries for help. 

The vIMMs showed participants not solely relying on papers' references but also looking at the context of the references of papers as provided by WoS~\cite{salisbury} or scite.

Contrasting previous evaluations of the used example system~\cite{schenqljournal}, we did not encounter design fixation~\cite{designfixation}. Even though participants were used to their individual workflows, they adapted to the possibilities of the example DL. When asked what users disliked or missed in the current system they did not appear to be stuck to the current design.


Improvements for our example system can function as markers to be incorporated into any DL which strives to better support its users: 
A system could include a shopping cart style option to save temporary results following Daffodil~\cite{daffodil}. Ranking possibilities for results, even soft ranking constraints and diversity aspects should be incorporated. Users should be suggested decompositions of their queries into parallel tasks and follow-up, likely queries to enable simple query specification~\cite{DBLP:conf/asist/Liu00B11}, e.g., recommend a query searching for most cited papers if a user has searched papers.

\subsubsection{Cost-Effectiveness}
The monetary cost of the proposed method is non-negligible: Users are required to take part in multiple sessions (in our case, 1.5h per person), which need to be transcribed. Nevertheless, this study type helps uncover discrepancies between users' typical task solution strategies and the functions of a DL with few study participants already.


\section{Conclusion}

To investigate our general research question \textit{How can we compare users' conceptions of search tasks in digital library with capabilities of such a system?}, we defined five more fine-grained research questions which we observed through a user study
with thirteen participants on two everyday tasks in digital libraries: Expert search and relevant paper search. 
We formalised study participants' usual task conduction processes in BPMNs, the ideal translation of these general processes to an exemplary DL system (SchenQL), and persons actually using the system for a qualitative DL evaluation.

Our experiment found the models to be beneficial and suitable for systematic qualitative DL evaluations. We were able to answer our research questions which investigated different aspects of users' ideal and actual task conduction (RQ$_1$, RQ$_2$), limitations of systems (RQ$_3$), a system's suitability for advanced tasks (RQ$_4$) and discovered discrepancies between users' perceptions and capabilities of systems (RQ$_5$).
In detail: (1) Users were willing to switch working spheres by explicitly using multiple tools to solve everyday tasks in DLs with their usual exploration strategy. 
(2) We observed users heavily relying on usage examples for the unknown DL and searching for papers on their chosen topic as entry point for their exploration tasks. 
(3) DL systems such as SchenQL cannot model interaction with experts, the incorporation of multiple data sources or the type of broad, explorative search as seen with general search engines. 
(4) Some study participants were able to complete the tasks using the unknown system by translating or adapting their ideal conduction models. 
(5) Participants tend to overmodel their usual task execution strategy and focus on a portion of the solution process while using SchenQL. 

We publish a reusable dataset (see Zenodo~\cite{DATASET}) resulting from our evaluation with transcripts of user interviews and the BPMN models, which would be classified as level-4 according to the 5-level system of Gäde et al.~\cite{gade_manifesto_2021}. 
%
It can be reused to construct new DL systems with requirements from vIMMs or viewing the vIMMs as a qualitative benchmark to investigate systems' compatibility with real user needs. 
New researchers can apply vIMMs as a starting point for deriving their individual information-seeking strategies. 
Additionally, one could try to use the BPMNs to construct usage models of different types of users of DLs with the prospect of formulating more real and diverse user types in simulation studies. One could also compare the BPMNs to existing search stratagems~\cite{Bates1989TheDO,DBLP:journals/scientometrics/KacemM18}.

Future work could apply our proposed qualitative evaluation technique to other DL search systems and combine these results in a meta-study. For prospective studies, eye-tracking software could be used to try to better explain user behaviour and annotate the process models. Different formal representations of user models could be incorporated and compared.

When focusing more on the results of our exemplary evaluation, users' readiness to switch systems could be investigated. 
Investigating potential links between study participants' expertise with DLs or their chosen topics and their task conduction could shed light on their individual cognitive loads~\cite{DBLP:journals/cogsci/Sweller88}. 



\bibliographystyle{IEEEtran}


\end{document}